\documentclass[prl,superscriptaddress,twocolumn,nofootinbib]{revtex4}
\usepackage{amsfonts,amssymb,amsmath}
\usepackage{mathrsfs}
\usepackage{color}
\usepackage{graphicx}
\usepackage{dcolumn}
\usepackage{bm}

\newcommand{\C}[1]{{\mathcal #1}}

\newcommand{\beq}{\begin{equation}}
\newcommand{\eeq}{\end{equation}}
\newcommand{\bea}{\begin{eqnarray}}
\newcommand{\eea}{\end{eqnarray}}

\newcommand{\half}{\frac 12}

\newcommand{\quarter}{\frac 14}

\newcommand{\Slash}[1]{{\ooalign{\hfil#1\hfil\crcr\raise.167ex\hbox{/}}}}

\begin{document}

\title{Reheating of the Universe as holographic thermalization}
\author{Shinsuke Kawai}
\email{kawai@skku.edu}
\affiliation{Department of Physics, Sungkyunkwan University,
Suwon 16419, Republic of Korea}
\author{Yu Nakayama}
\email{nakayama@theory.caltech.edu}
\affiliation{California Institute of Technology, 452-48, Pasadena, California 91125, USA}
\affiliation{Kavli Institute for the Physics and Mathematics of the Universe (WPI), Todai Institutes for Advanced Study, Kashiwa, Chiba 277-8583, Japan}
\date{September 15, 2015}

\begin{abstract}
Assuming gauge/gravity correspondence we study reheating of the Universe using its holographic dual.
Inflaton decay and thermalisation of the decay products correspond to collapse of a spherical shell and formation of a blackhole in the dual anti-de Sitter (AdS) spacetime. The reheating temperature is computed as the Hawking temperature of the developed blackhole probed by a dynamical boundary, and is determined by the inflaton energy density and the AdS radius, with corrections from the dynamics of the shell collapse. 
For given initial energy density of the inflaton field the holographic model gives significantly lower reheating temperature than the instant reheating scenario, while it is shown to be safely within phenomenological bounds.
\end{abstract}

\maketitle

%
According to the standard lore of inflationary cosmology, reheating of the Universe is caused by out-of-equilibrium decay of the inflaton field that oscillates about its potential minimum.
Although reheating is a crucial process that determines subsequent thermal history of the Universe,
our understanding of it is still incomplete 
as the decay process down to the Standard Model (SM) particles is highly involved.
There are several phenomenological models of reheating, each giving different evaluation of the reheating temperature.
Among these, the most traditional one is due to perturbative Born decay of the inflaton, in which the reheating temperature is computed from the condition that the inflaton decay rate $\Gamma$ becomes comparable to the Hubble expansion rate $H$, as
\begin{align}\label{eqn:Tprh}
T_{\rm prh}\approx \left(\frac{90}{\pi^2 g_*}\right)^{\frac{1}{4}}\left(M_{\rm P}\Gamma\right)^{\frac 12}.
\end{align}
Here, $g_*$ is the relativistic degrees of freedom at the time of reheating,
$M_{\rm P}\equiv (8\pi G_4)^{-1/2}=2.4\times 10^{18}$ GeV is the reduced Planck mass and $G_4$ is
the four-dimensional Newton constant.
This Born decay picture is known to be too simplistic, at least in some cases, as nonperturbative resonance effects can change the decay rate drastically. 
In the scenario of {\em preheating} \cite{Kofman:1994rk,Kofman:1995fi,Kofman:1997yn}, reheating is assumed to take place in three steps: the nonperturbative resonant decay of the inflaton, followed by perturbative cascade decay of the decay products, and then eventual thermalisation.
There exist proposals of other reheating mechanisms, including those based on evaporation of primordial blackholes \cite{Barrow:1990he,GarciaBellido:1996qt,Hidalgo:2011fj}, surface evaporation of Q-balls \cite{Enqvist:2002rj}, and nonminimal gravitational coupling of the inflaton \cite{Bassett:1997az}. 
We discuss, in this Letter, a novel description of reheating based on the gauge/gravity duality \cite{Maldacena:1997re,Gubser:1998bc,Witten:1998qj,Witten:1998zw}.
This may be considered as the limit opposite to the perturbative scenario and is supposed to take into account strongly coupled dynamics of the thermalisation process.

Following the idea of holographic thermalisation 
\cite{Witten:1998zw,Danielsson:1999zt,Danielsson:1999fa,Bhattacharyya:2009uu,Balasubramanian:2010ce,Balasubramanian:2011ur} which asserts that blackhole formation in a $(d+1)$-dimensional anti-de Sitter (AdS) spacetime is a dual description of out-of-equilibrium thermalisation in $d$-dimensional conformal field theory (CFT), we postulate that the Universe sits at the boundary of a five-dimensional asymptotically AdS spacetime.
We shall consider, schematically, the boundary action of the form
\begin{align}\label{eqn:Sbdry}
S_{\rm bdry}=S_{\rm CFT}+\int d^4x \Phi_0(\tau){\C O}(\tau),
\end{align}
and regard $S_{\rm CFT}$ as the action of the Universe including (but not restricted to) the SM matter.
Here we treat the inflaton as an external field that is not included in the matter of the Universe.
The operator $\Phi_0(\tau)$ represents the oscillating inflaton and ${\C O}(\tau)$ is the matter in the Universe that couples to the inflaton\footnote{
In the two-body scattering into two bosons $\phi\phi\to\chi\chi$, for example, $\Phi_0=\phi^2$ and ${\C O} =\chi^2$. 
In the case of Higgs inflation the Higgs field ought to be split into the massive (inflaton) part $\Phi_0$ and the nearly massless (SM) part which is in the CFT.
}.
Aside from the interaction with the inflaton, the matter content of the Universe is nearly massless
at high energies and may be modelled as a CFT.
Prior to reheating the Universe must have undergone a rapid adiabatic expansion, i.e. inflation.
Therefore the CFT is at zero-temperature when reheating commences.
Our use of holography is motivated by the success of holographic quantum chromodynamics (QCD)
\cite{CasalderreySolana:2011us};
the energy scale of reheating may well be higher than that of the quark-hadron phase transition, and then the ``radiation" in the Universe should be composed of ultra-relativistic 
quark-gluon plasma.
We will not, nevertheless, specify the particle content of the CFT in the discussion below.
Although a legitimate use of gravity dual would certainly require a large number of colours $N$, we will take a phenomenological approach and assume the existence of the gravity dual. 
Our focus here is on what the gravity dual will tell us about reheating of the Universe.

The out-of-equilibrium decay of the inflaton is a process of transferring its energy to the matter in the Universe.
This can be seen as disturbance of the CFT by an external shock which is the
oscillating inflaton operator $\Phi_0(\tau)$ in \eqref{eqn:Sbdry}.
The time scale of the disturbance $\Delta\tau$ may be determined by the decay efficiency and Hubble damping.
In the gravity dual, the thermalisation corresponds to formation of a blackhole in AdS${}_5$, 
caused by collapse of a shell that destabilises the pure AdS.
The thickness of the shell corresponds to the time scale of reheating $\Delta\tau$.
The boundary conditions of the infalling shell should be given by the oscillating field $\Phi_0(\tau)$ of the boundary action \eqref{eqn:Sbdry}, in accordance with the GKPW prescription
\cite{Maldacena:1997re,Gubser:1998bc,Witten:1998qj,Witten:1998zw}.

The dynamics of blackhole formation in the asymptotically AdS spacetime is described by the AdS-Vaidya solution \cite{Wang:1998qx},
\begin{align}\label{eqn:Vaidya}
ds^2 =& -f(r,v) dv^2 +2dv dr +r^2 d\Omega_3^2,\cr
f(r,v) =& 1 +\frac{r^2}{L^2} -\frac{r_0^2}{r^2}\theta(v),
\end{align}
where $L$ is the AdS radius and $r_0$ is related to the mass of the five-dimensional blackhole by
\begin{align}\label{eqn:BHmass}
M_5=\frac{3\pi r_0^2}{8G_5}.
\end{align}
Here, $G_5$ is the five-dimensional Newton constant.
The function $\theta$ asymptotes to $\theta\to 0$ inside the shell and
$\theta\to 1$ outside, and thus the AdS-Vaidya solution interpolates the pure AdS solution in the past (inside the shell) and the AdS-Schwarzschild solution in the future (outside).
With the change of variables
$dv=dt+f(r,v)^{-1} dr$, the metric in the static coordinates reads
\begin{align}
ds^2 = -f(r) dt^2+\frac{dr^2}{f(r)} +r^2 d\Omega_3^2,
\end{align}
in which the function $f(r)$ behaves as
\begin{align}
f(r)\to 
\left\{\begin{array}{ll}
f_-(r) \equiv 1+\frac{r^2}{L^2} &(\text{inside}), \cr
f_+(r) \equiv 1+\frac{r^2}{L^2}-\frac{r_0^2}{r^2} &(\text{outside}).\end{array}\right.
\end{align}
After the shock passes, the metric seen by a local observer becomes AdS-Schwarzschild, indicating that the CFT at the boundary is thermalised.
The temperature of the CFT will be given by the Hawking temperature of the AdS-Schwarzschild blackhole, which may be interpreted as the reheating temperature of the Universe.

\begin{figure}[t]
\includegraphics[width=65mm]{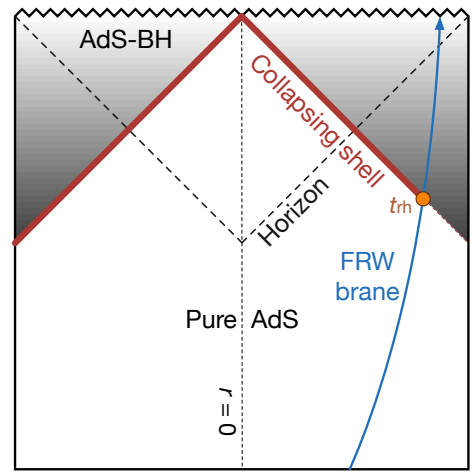}
\caption{\label{fig:Penrose}
The Penrose diagram of the AdS-Vaidya solution describing blackhole formation.
The Universe is considered as a probe {\em FRW brane}, a hypersurface 
solving Israel's junction conditions.
The collapsing shell is released form the brane during reheating, with boundary conditions given by $\Phi_0(\tau)$.
The region right to the FRW brane is to be excised so that the brane may represent a true boundary of the spacetime.
}
\end{figure}

Cosmological application of holography has been actively studied since the early days of AdS/CFT correspondence. 
If we are to consider the Friedman-Robertson-Walker (FRW) universe as the CFT side of the correspondence, we are faced with two apparent obstacles. 
An expanding universe is a weakly gravitating system and hence the boundary theory in such a setup is not entirely decoupled from gravity.
The other issue is the time dependence of the temperature;
in contrast to the standard GKPW prescription for flat space CFT in which the overall scaling of the temperature is unfixed, the temperature in cosmology has a definite value and redshifts as the inverse of the scale factor, $T\propto a^{-1}$.
These features suggest that when discussing cosmology in AdS/CFT, the boundary theory should be treated dynamical \cite{Gubser:1999vj}.
The Universe is then envisaged as a hypersurface moving in the asymptotically AdS${}_5$ bulk.

To proceed, we make use of the observation 
\cite{Garriga:1999bq,Verlinde:2000wg,Savonije:2001nd,
Chamblin:1999ya,Chamblin:2000bx} that the Friedman equation is obtained from the induced metric on a hypersurface 
in a five-dimensional AdS-Schwarzschild (or AdS-Vaidya) spacetime.
The emerging Friedman equation is
\begin{align}\label{eqn:Friedman}
H^2=-\frac{1}{a^2}+\frac{r_0^2}{a^4}+\cdots,
\end{align}
where the ellipses represent terms that come from extra matter on the brane, which are not important in our discussion of reheating and will be neglected.
The second term in \eqref{eqn:Friedman} is a radiation-like contribution proportional to the mass of the 
five-dimensional blackhole.
In many of the brane universe literature this term is treated as an extra contribution {\em in addition to} the matter of the Universe, but here in holographic reheating this term is naturally interpreted as the thermal radiation resulting from thermalisation of the shock.
The first term of \eqref{eqn:Friedman} is the curvature term $-k/a^2$, indicating that we must consider the closed ($k=1$) FRW universe.

The identification of the FRW metric and the induced metric on the hypersurface implies that the scale factor of the universe is the AdS radial coordinate, $a=r$.
An expanding universe is thus a brane moving away from the centre of the AdS.
Fig.\ref{fig:Penrose} shows embedding of the FRW universe in the AdS-Vaidya spacetime.
Reheating takes place at the transition from the pure AdS to the AdS-Schwarzschild background, marked by the small orange circle.
As we regard the FRW brane to be a true boundary of the spacetime, the region to the right of the brane is understood to be cut away.
The collapsing shell is released from the brane with the boundary conditions given by $\Phi_0(\tau)$.
The horizon develops as the shell collapses.
Its location $r=r_+$ is found as a solution to $f_+(r)=0$,
\begin{align}\label{eqn:r+}
r_+ \equiv
L\left[\half \left(
\sqrt{1+\frac{4r_0^2}{L^2}}-1
\right)
\right]^\half.
\end{align}
The Hawking temperature of the blackhole seen by a static observer is computed in the standard way by Euclideanising the near-horizon metric.
The absence of conical singularity then gives
\begin{align}\label{eqn:Tstat}
T_{\rm stat}(r)=\frac{2r_+^2+L^2}{2\pi L^2 r_+}\frac{1}{\sqrt{f_+(r)}},
\end{align}
where the factor $1/\sqrt{f_+(r)}$ is due to gravitational redshift (Ehrenfest-Tolman effect).
This $T_{\rm stat}$ cannot represent the temperature on the probe brane as it is ill-behaved near the horizon.
Nevertheless, it is expected to coincide with the temperature of the probe when the probe brane is far outside the horizon and nearly static.
Thermodynamics on the brane suggests that the natural time scale on the moving brane is 
$d\tau=\frac{a}{L}dt$ \cite{Savonije:2001nd}, from which the temperature of the probe brane is found to be
\begin{align}\label{eqn:Tprobe}
T_{\rm probe}=\frac{2r_+^2+L^2}{2\pi L r_+}\frac{1}{a}.
\end{align}
This is regular at the horizon and coincides with \eqref{eqn:Tstat} when $a\gg r_0$, $a\gg L$.
Hence \eqref{eqn:Tprobe} is qualified to be the temperature of the FRW universe.

Apart from the scale-factor dependent redshift, $T_{\rm probe}$ is determined by the five dimensional blackhole mass $M_5$ through \eqref{eqn:BHmass} and \eqref{eqn:r+}.
The mass of the blackhole, in turn, encodes the information of inflaton decay.
While detailed process of reheating may be very much involved, in the gravity dual at least energy conservation is expected.
As the blackhole results from the collapse of the shell, it is natural to assume
\begin{align*}
M_5
=\varepsilon\times(\text{area of shell})_*\times(\text{energy density of shell})_*
\end{align*}
where $\varepsilon$ is the efficiency ($0<\varepsilon\leq 1$) of blackhole formation and the asterisk denotes quantities evaluated at the onset of reheating.
The shell is spherical and its area is $2\pi^2 r_*^3=2\pi^2 a_*^3$.
The energy density of the shell is related to that of the oscillating inflaton $\rho_*$ at the onset of the collapse.
As we consider the closed universe, the total inflaton energy is finite and is given by
$2\pi^2a_*^3\rho_*$.
Including the redshift $a_*/L$ that converts energy on the brane to that in the shell frame\footnote{
The cosmic time on the brane before and after the transition may be different.
However, the difference is immaterial as our interest is only in the asymptotic region where $f_-(a)\approx a^2/L^2\approx f_+(a)$.
}, the blackhole mass is written as
\begin{align}\label{eqn:M5}
M_5=& \frac{2\pi^2 \varepsilon a_*^4 \rho_*}{L}.
\end{align}
Given a model of inflation, the inflaton energy density $\rho_*$ may be evaluated explicitly.
For an inflaton field $\varphi$ with mass $m$ and negligible self-interaction, for example,
\begin{align}\label{eqn:chaoticrho}
\rho_*=3M_{\rm P}^2H_*^2=\left[\half\dot\varphi^2+\half m^2\varphi^2\right]_*\approx m^2\varphi_*^2,
\end{align}
with $\varphi_*$ the initial amplitude of the oscillating inflaton.

To interpret \eqref{eqn:Tprobe} as the temperature of the Universe, it is important to note that the phase structure of the thermodynamics of an AdS-Schwarzschild blackhole is entirely different from the asymptotically flat Schwarzschild case.
When $r_0\ll L$, we find $T_{\rm probe}\approx{L}/{2\pi r_0a}$.
This phase is similar to the asymptotically flat case and exhibits instability due to negative specific heat; clearly, it does not represent the thermal equilibrium of the Universe.
Taking the opposite limit $r_0\gg L$, which is equivalent to choosing 
large $a_*$ (this is natural due to inflation before reheating), the temperature \eqref{eqn:Tprobe} behaves as
\begin{align}\label{eqn:Tdec}
T_{\rm probe}\approx\frac{r_0^\half}{\pi L^\half a}.
\end{align}
The specific heat is positive in this phase, appropriate for the reheating model.
This is the deconfinement phase in QCD.
Now using \eqref{eqn:BHmass} and \eqref{eqn:M5} in \eqref{eqn:Tdec} the temperature of the FRW universe may be expressed as
\begin{align}
T=\left(\frac{8G_5 M_5}{3\pi^5L^2}
\right)^\quarter\frac{1}{a}
=\left(\frac{16 G_5\varepsilon\rho_*}{3\pi^3L^3}
\right)^\quarter\frac{a_*}{a}.
\end{align}
Denoting the scale factor at the moment of thermalisation (end of reheating) as $a_{\rm rh}$, the reheating temperature of the Universe is written as
\begin{align}\label{eqn:Thrh1}
T_{\rm hrh}=\left(\frac{16 G_5\varepsilon\rho_*}{3\pi^3L^3}
\right)^\quarter\frac{a_*}{a_{\rm rh}}.
\end{align}
This temperature may be expressed using the relation between the four- and five-dimensional Newton constants $G_5=G_4 L/2=L/16\pi M_{\rm P}^2$ 
\cite{Gubser:1999vj} 
as
\begin{align}\label{eqn:Thrh2}
T_{\rm hrh}=\left(\frac{\varepsilon\rho_*}{3 M_{\rm P}^2L^2}
\right)^\quarter\frac{a_*}{\pi a_{\rm rh}}.
\end{align}
Thus, in the holographic model the reheating temperature is determined by the inflaton energy density $\rho_*$, the AdS radius (characterising the CFT) $L$, as well as by the efficiency of the blackhole formation $\varepsilon$ and the redshift during reheating $a_*/a_{\rm rh}$. 
The efficiency $\varepsilon$ depends on details of the collapsing dynamics 
and may be evaluated numerically.
The redshift $a_*/a_{\rm rh}$ is related to the function $\theta(v)$ of \eqref{eqn:Vaidya} that determines the thickness of the shell.
As most of the shell energy is expected to be used in blackhole formation and the Hubble expansion during reheating is not large, it is natural to suppose that both $\varepsilon$ and $a_*/a_{\rm rh}$ are not much smaller than ${\C O}(1)$.

To illustrate our results, let us compare our scenario with instant reheating in which the inflaton energy density $\rho_*$ is instantly converted into the energy density of radiation. 
The temperature of instant reheating follows from the Stefan-Boltzmann law and reads
\begin{align}\label{eqn:Tirh}
T_{\rm irh}=\left(\frac{30\rho_*}{\pi^2 g_*}\right)^\quarter.
\end{align}
Comparing \eqref{eqn:Thrh1} and \eqref{eqn:Tirh}, we may define the effective degrees of freedom for holographic reheating,
\begin{align}\label{eqn:geff}
g_*^{\rm eff}
=&\frac{45\pi L^3}{8G_5\varepsilon}\left(\frac{a_{\rm rh}}{a_*}\right)^4.
\end{align}
The $L^3$ dependence is expected from the central charge $c\sim N^2\sim L^3/G_5$
of strongly coupled 4-dimensional CFT in the deconfinement phase.

As an example, let us consider the $m^2\varphi^2$ chaotic inflation model with the Planck-normalised inflaton mass $m=1.5 \times 10^{13}$ GeV.
The amplitude of the oscillating inflaton is $\varphi_*\approx \sqrt 2 M_{\rm P}$ at the end of slow roll and the inflaton energy density \eqref{eqn:chaoticrho} is found as $\rho_*\approx 8\times 10^{-11}M_{\rm P}^4$.
Using the relativistic degrees of freedom $g_*^{\rm SM}\sim 100$ of the SM, the instant reheating scenario yields somewhat high
reheating temperature $T_{\rm irh}\sim 3\times 10^{15}$ GeV.
In the holographic scenario, the effective degrees of freedom \eqref{eqn:geff} may be written using \eqref{eqn:Thrh2} as
\begin{align}
g_*^{\rm eff}
=\frac{90\pi^2M_{\rm P}^2L^2}{\varepsilon}
\left(\frac{a_{\rm rh}}{a_*}\right)^4.
\end{align}
Since $\varepsilon\leq 1$ and $a_{\rm rh}>a_*$, and as the AdS radius should be larger than the Planck scale $M_{\rm P} L\gtrsim 1$, 
the holographic effective degrees of freedom $g_*^{\rm eff}$ must be larger than $g_*^{\rm SM}$.
Correspondingly, the holographic reheating temperature $T_{\rm hrh}$ is lower than 
$T_{\rm irh}$, given the same energy density of the decaying inflaton.
How large $g_*^{\rm eff}$ can be?
The nucleosynthesis bound of the reheating temperature is $T_{\rm rh}\gtrsim$
a few MeV \cite{DeBernardis:2008zz,Hannestad:2004px,Kawasaki:2000en,Kawasaki:1999na}.
However, for the holographic model it is more appropriate to take the quark-hadron phase transition $\sim 200$ MeV as the lower bound of the temperature of the strongly coupled CFT,
which gives 
$g_*^{\rm eff}\lesssim 5\times 10^{66}$.
Thus sufficiently large parameter space is left for the scenario of holographic reheating.

We have discussed post-inflationary reheating scenario based on holographic thermalisation.
The reheating temperature is given by \eqref{eqn:Thrh2}, which is to be compared with the perturbative decay scenario \eqref{eqn:Tprh} or the instant reheating scenario \eqref{eqn:Tirh}.
The strongly coupled CFT which models the particle theory of the Universe is characterised by the AdS radius $L$, and the gauge coupling and the number of colours $N$ are encoded in it.
This scenario certainly has limited validity and is expected to be useful only when strongly coupled dynamics dominates. 
There are many issues to be investigated further; to conclude, we comment on some of them.
The shell collapse picture we have used is consistent only for the closed FRW universe.
Nevertheless, the expression for the reheating temperature \eqref{eqn:Thrh2} is independent of the curvature radius and thus may be applicable to the flat universe as well.
Extending the holographic model to the open FRW, however, may have difficulties related to topological issues \cite{Witten:1999xp}.
It would be also interesting to see if our scenario can accommodate lepto/baryogenesis, presumably by including vector degrees of freedom in the AdS.
Another topic we have not discussed is string theoretical construction. While certainly important, finding a concrete D-brane configuration would require, at least partially,
dynamical treatment.
Finally, inhomogeneity of the decay process can be important since, for example, modulated reheating may create cosmological structures \cite{Kofman:2003nx,Dvali:2003em}.
Recent numerical study has uncovered rich structure of blackhole formation in AdS background 
\cite{Bizon:2011gg,Jalmuzna:2011qw,Maliborski:2013jca,Bizon:2015pfa},
which may find cosmological applications in the physics of reheating, in particular, of modulated reheating.


\bigskip

{\em Acknowledgements.}---
We acknowledge helpful conversations with A. Rostworowski.
This work was supported in part by 
Grant-in-Aid for Scientific Research NRF-2012R1A1A2007575 (S.K.),
a Sherman Fairchild Senior Research Fellowship
at California Institute of Technology and 
DOE grant number de-sc0011632 as well as 
WPI Initiative, MEXT (Y.N.).
\bigskip



\end{document}